\documentclass[aps,twocolumn,floats,showpacs]{revtex4}
\usepackage{graphicx}

\begin{document}
\title{Novel universality class of absorbing transitions with 
continuously varying critical exponents}
\author{Jae Dong Noh}
\affiliation{Theoretische Physik,
Universit\"{a}t des Saarlandes, 66041 Saarbr\"{u}cken, Germany}
\author{Hyunggyu Park}
\affiliation{School of Physics, Korea Institute for Advanced Study,
         Seoul 130-012, Korea}
\date{\today}

\begin{abstract}
The well-established universality classes of absorbing critical phenomena
are directed percolation (DP) and directed Ising (DI) classes.
Recently, the pair contact process with diffusion (PCPD) has been 
investigated extensively and claimed to exhibit a new type of critical 
phenomena distinct from both DP and DI classes. 
Noticing that the PCPD possesses a long-term memory effect,
we introduce a generalized version of the PCPD (GPCPD) with
a parameter controlling the memory effect. 
The GPCPD connects the DP fixed point to the PCPD point continuously.
Monte Carlo simulations show that
the GPCPD  displays novel type critical
phenomena which are characterized by continuously varying critical exponents.
The same critical behaviors are also observed in models where two
species of particles are coupled cyclically.
We suggest that the long-term memory may serve as
a marginal perturbation to the ordinary DP fixed point.
\end{abstract}


\maketitle

\section{Introduction}\label{sec:intro}
A nonequilibrium system with trapped~(absorbing) states may display
a so-called absorbing phase transition between an inactive and an active
phase~\cite{Marro,Haye}. A state which has a zero transition
probability into any other state is called the absorbing state.
A system in the inactive phase always evolves into the absorbing 
state and stays there forever. On the other hand, a system in the active 
phase may not be trapped in the absorbing state with a finite probability.
There has been growing interest in the critical behaviors of the
absorbing phase transitions since a wide range of phenomena such as
epidemic spreading, catalytic chemical reactions, and surface
roughening, display absorbing transitions~\cite{Marro,Haye}.

Besides their wide applications, absorbing critical phenomena have been
the focus of a number of theoretical works, since they are categorized into a few 
universality classes characterized by the symmetry between 
the absorbing states and/or the conservation in 
dynamics~\cite{GKT,PP,Hinrichsen97,Hwang98}.
Criticality of each universality class is described by three independent
critical exponents; $\beta$ for the order parameter, $\nu_\perp$ for the
correlation length, and $\nu_\parallel$ for the relaxation time.
For systems which are free from quenched disorder and evolve only through 
short-range processes, the directed percolation~(DP) and the directed Ising~(DI)~(or
parity conserving) classes are well-established ones.

The DP class involves typically a single absorbing state without
any kind of conservation in dynamics~\cite{Janssen81,Grassberger82}.
The contact process, a model for epidemic spreading, is 
a prototypical example of the DP class~\cite{Harris}. 
In this model, individuals on a lattice are either infected or healthy.
Infected ones may be healed spontaneously or infect healthy neighbors.
There is a single absorbing state where all individuals are healthy. 
Varying the relative rates between infection and healing processes, 
one can find a phase transition from the absorbing phase into 
the active phase. 

Its stationary and dynamic critical behaviors are characterized by
spatiotemporal cluster patterns of infected individuals. 
These clusters can be mapped to the directed percolation clusters~\cite{Marro},
when the temporal direction is set to be the preferred direction of DP clusters.
So all critical exponents (fractal dimensions) take the same values as
the corresponding DP critical exponents.
Most of systems with absorbing states belong to this DP class, e.g.~the 
Domany-Kinzel cellular automaton~\cite{Domany-Kinzel}, 
the Ziff-Gulari-Barshad model for a
surface catalytic reaction~\cite{ZGB}, the branching-annihilating random
walks with an odd number of offspring~\cite{BAW}, and the pair contact
process~(PCP)~\cite{PCP}. Unlike others, the PCP has infinitely many
absorbing states, which leads to controversial transient behaviors, i.e.,
nonuniversal scaling~\cite{PCP,PCP2,Mendes,DbA} versus absence of 
scaling~\cite{GCR,DH}. But its stationary critical 
behavior still belongs to the DP class at low dimensions~\cite{Wijland02}. 

The DI class includes systems with two equivalent 
absorbing states with Ising-like $Z_2$ symmetry or equivalently 
in one dimension (1D)  a single absorbing state with parity conservation 
in the domain wall language. 
The nonequilibrium kinetic Ising (NKI) model with combined 
zero-temperature spin-flip dynamics and infinite-temperature spin-exchange 
dynamics is an example of the DI systems~\cite{Menyhard}. 
In this model, only Ising spins near domain walls
can flip, so that the two states with all spins up or down are absorbing.
These two absorbing states are probabilistically equivalent.
In terms of the domain wall, there is a single absorbing state (vacuum)
with parity conservation in the number of domain walls, since spin
flips change it only in pairs. 
Other examples in the DI class includes the interacting monomer-dimer
model~\cite{IMD}, the branching-annihilating random walks with an even number of
offspring~\cite{BAW,Jensen94}, and generalized versions of the contact
process~\cite{Hinrichsen97,Inui}. There also exist models in
the DI universality class that have infinitely many absorbing
states~\cite{Hwang_Park,Marques99,HSPark}. 
They also display non-universal scaling behaviors in the transient 
regime~\cite{Marques99,HSPark}. 

There is an {\em infinite dynamic barrier} between
two absorbing states of DI systems, which is similar to the free energy 
barrier between two ground states in the ordered phase of the equilibrium 
Ising system~\cite{Hwang_Park}. 
A state near one absorbing state cannot evolve into a state near the other
absorbing state by a finite number of successive local changes.
In other words, a frustration (domain wall)
in a configuration generated by pasting two absorbing configurations
cannot disappear within a finite number of time steps.
For example, a spin state in the 1D NKI model with all
spins up in one semi-infinite lattice and all spins down in the other
semi-infinite lattice, never relaxes to the absorbing state.

The concept of the infinite dynamic barrier is very useful to understand
the critical behavior of systems with infinitely many absorbing
states. For instance, the PCP~\cite{PCP} and the modified interacting
monomer-dimer~(IMD-IMA)
model~\cite{Hwang_Park} have infinitely many absorbing states. In the PCP,
a frustration between any of the absorbing states can disappear locally, so
the absorbing transition falls into the DP class. However, in the IMD-IMA, 
the infinite dynamic barrier separates the absorbing states into two equivalent
groups of absorbing states, which results in the DI-type critical behavior~\cite{Hwang_Park}.

A few exceptional cases have been reported: L\'evy-type long-range flights
are relevant to the absorbing critical phenomena, which lead to continuous
variation of critical exponents~\cite{Jan_Hin}. Multi-species particle
reaction-diffusion systems also show non-DP and non-DI critical
behaviors, where an inter-species hardcore interaction plays
an important role~\cite{Kwon}. A lattice gas model with 
a global conservation in the particle number shows a  novel type
criticality related to self-organized critical systems~\cite{Rossi}. 
The nonequilibrium $q$-state Potts models in higher dimensions show 
interesting critical behaviors~\cite{Lipowski}.
A quenched randomness also leads to novel absorbing critical
phenomena~\cite{Qrandom}.

Recently, Howard and T\"{a}uber introduced a modified PCP~(called
as PCPD) model, which allows single particle 
diffusion~\cite{Howard_Tauber,addG}. 
They studied the PCPD in the context of bosonic
field theory and showed that the field theory is non-renormalizable
and the absorbing transition does not belong to the DP class. 
A fermionic version was first studied by Carlon, Henkel, and 
Schollw\"{o}ck~\cite{Carlon} in 1D, 
which raises continuing debate on the universality 
class~\cite{Carlon,Haye2,Odor1,CGrass,Haye3,KPark,Odor2}.
Hereafter, we only focus on the 1D systems.

The PCPD is defined on a lattice, each site of which is either occupied
by a particle~($X$) or empty~($\emptyset$). Dynamic rules are
given as follows. A nearest neighbor pair of particles can either
annihilate~($XX\rightarrow \emptyset\emptyset$) with probability
$p(1-d)$ or branch one offspring~($XX\rightarrow XXX$) to one of
the neighboring sites with probability $(1-p)(1-d)$. A particle
can hop to its neighboring site with probability $d$.
Branching and hopping attempts are rejected if a particle would
land on the top of another particle. 

When the diffusion is not allowed ($d=0$), it reduces to the ordinary 
PCP with infinitely many absorbing states. Any state with only isolated particles 
is an absorbing state. But there is no infinite dynamic barrier between them
and the transition belongs to the DP class~\cite{PCP}. 

At nonzero $d$, the PCPD has only {\em two} absorbing states; a vacuum
state and a state with a single diffusing particle.
In fact, the latter forms an absorbing subspace consisting of $L$ configurations 
with system size $L$ in 1D. 
Once the system evolves into this absorbing subspace,
it drifts freely inside the subspace but cannot escape out of it. 
The threshold transfer process~\cite{Mendes} is one example  
with an absorbing subspace. But it does not contain any other absorbing state
and displays the DP-type transition. 

The structure of absorbing states in the PCPD is
unique with one point-like absorbing state and one absorbing subspace.  
It is clear that  there is no infinite dynamic barrier between these absorbing states.
Hence, one may argue that the PCPD should belong to  
the DP class.
However, this argument turns out to be premature.
Background diffusing solitary particles 
generate a {\em long-term memory effect} on the order 
parameter (pair density). Solitary particles detached from different trains
of particles diffuse and collide each other to create a new
particle pair, which leads to history dependence in the pair-creation
rate. This process is governed by annihilating random
walks, where the colliding probability of two walkers decays
algebraically with time. This long-term memory effect might be relevant 
and lead to a new type of critical phenomena.

A numerical investigation using  density matrix renormalization group 
techniques~\cite{Carlon}
revealed that numerical values of some critical exponent ratios are 
close to the DI values rather than the DP values. 
Subsequent extensive simulations~\cite{Haye2,Odor1,CGrass} seem to exclude
the possibility of both DI and DP classes and suggest that the PCPD belongs
to a new universality class. 
However, huge corrections to scaling conceal the true 
asymptotic scaling behavior and numerical estimates for the critical exponents 
are obscure with considerable uncertainty.
Similar critical behaviors were also observed 
in related models~\cite{Haye3,KPark,Odor2}, which include
a modified PCPD~\cite{KPark} with the branching process $2X\rightarrow 4X$ 
instead of $2X\rightarrow 3X$ and an effective model with two species of
particles coupled cyclically~\cite{Haye3}. 

Despite all those efforts, universal features that could characterize the novel 
universality class were not uncovered yet. We explore this issue in the
present work. Hinrichsen noticed that two types of degrees of freedom are 
present in the PCPD; a particle pair that can branch and annihilate, and 
a diffusing solitary particle~\cite{Haye3}.
All activities that can change the number of particles are carried 
out by the particle pairs. So the particle pair density can be regarded as 
an order parameter. The two degrees of freedom are coupled cyclically; 
one can be transmuted to the other, and vice versa, via  particle diffusion. 
The cyclic coupling results in the long-term memory effect as described above, 
which we believe plays a crucial role in this novel critical phenomena. 

These observations
lead us to consider a generalized PCPD~(GPCPD) model with 
a parameter controlling the strength of the memory effect, which will be
explained in Sec.~\ref{sec:model}. Without the memory effect, 
the GPCPD should fall into the DP class. So, our parameter
connects the DP fixed point to the PCPD point, which allows us
to study the long-term memory effect on the DP-type models systematically.

One may expect that the long-term memory would serve as a relevant 
(at least, marginal) perturbation to the DP fixed point, in order to
account for the non-DP type novel critical behavior at the PCPD point.
In that case, one can hope that, by varying the control parameter,
those huge corrections to scaling found at the PCPD point might be 
reduced to a tolerable level, so the moderate numerical efforts may
reveal the true asymptotic behavior. 

We performed extensive Monte Carlo simulations in Sec.~\ref{sec:MC}.
Our numerical results show that corrections to scaling 
are huge only near the PCPD point and the asymptotic regimes 
are rather easily reached at other values of the control parameter.
So, we were able to estimate the values of the critical exponents 
with reasonable accuracy for a wide range of the control parameter,
except not very near the PCPD point. In the absence of the memory effect,
we find the DP universality class as expected. 
With the memory effect, surprisingly, 
we observe continuous variation of critical
exponents with the memory strength. Especially, the order parameter
exponent $\beta$ varies more than 60\%, which is  far beyond 
the statistical errors less than at most 10\%. 

In order to establish this novel universality class with continuously varying 
critical exponents firmly, in Sec.~\ref{sec:universality}, 
we introduce two independent model systems with
two species of particles: one species plays the role of the particle pairs and 
the other the role of the diffusing solitary particles. They are coupled cyclically
through transmutations, which leads to the long-term memory effect.
These models are also studied by Monte Carlo simulations, and are shown
to display the same type of critical phenomena as in the GPCPD.

In Sec.~\ref{sec:lifetime}, we suggest
one possible scenario to account for this novel universality class
with continuously varying exponents. This scenario assumes that the long-term memory
plays the role of the marginal perturbation to the DP fixed point. 
We numerically check this scenario by measuring the life time distribution of
solitary particles. Our analysis shows that this scenario may be considered 
as a reasonable one, though it needs full field-theoretical analysis for
a definitive test. 
Finally, we summarize our works in Sec.~\ref{sec:sum&dis}.

\section{Model}\label{sec:model}
The GPCPD model is defined on a 1D lattice of size $L$ with the periodic boundary
conditions. Each site is either occupied by a particle~$(X)$ or empty~$(\emptyset)$. 
The system evolves in
time according to the dynamic rules of the PCPD~(see Sec.~\ref{sec:intro}) 
with one additional ingredient: When the hopping of a solitary particle creates 
a new particle pair through colliding with another particle, 
this new particle pair annihilates instantaneously with probability 
$1-r$ or survives with probability $r$.  
This does not apply to particle pairs formed  by branching processes.
At $r=0$, two solitary particles always
annihilate upon meeting and have no chance to turn into a nearest neighbor particle pair.
The parameter $r$ controls the transmutation rate of
two solitary particles into a nearest neighbor particle pair upon
meeting. At $r=1$, the model reduces to the ordinary PCPD.

As in the PCP, we take the particle pair density as the order parameter.
At nonzero $r$, there is a feedback mechanism to
increase the pair density via the collision of diffusing 
solitary particles. A solitary particle is created from
a train of particles by a hopping (detaching) process of 
a boundary particle or a pair annihilation process of a triplet of particles. 
In this process, at least one particle pair
is sacrificed. The solitary particles diffuse and collide with
each other to form  a particle pair with probability $r$. 
So, one can say that particle pairs temporarily turn into 
solitary particles and resurrect later by collision of those solitary particles.
The colliding probability of two diffusing particles usually decays algebraically in time.
Therefore, this feedback mechanism induces long-term history dependence,
which is called the {\em long-term memory effect}, of the order parameter.
As $r$ increases, the long-term memory effect becomes stronger.
The PCPD model ($r=1$) has the maximum memory effect.

The $r=0$ point is special. Collision of two particles does not generate 
a particle pair at all, so there is no feedback mechanism for the pair density.
Once the system evolves into a state without any particle pair, 
it stays inside a subspace of states without a pair
and, in the end, will be trapped into one of the absorbing states 
(the vacuum state and the states with a single particle).
So, in the viewpoint of the order parameter as the pair density,
the no-pair subspace serves as the absorbing subspace 
which  the system cannot escape from. Evolution inside this subspace is 
governed by the trivial pair annihilation dynamics of diffusing particles
that have no memory effects on the order parameter fluctuations. Therefore,
the GPCPD at $r=0$ should be effectively equivalent to the PCP model without
diffusion. The no-pair subspace is exactly the same as the absorbing subspace 
of the latter. The coupling constants for the branching and annihilation processes
are renormalized due to the inclusion of the single-particle diffusion, and 
the critical point is shifted. However, the critical fluctuations should be
identical and we expect the DP-type stationary critical behavior at the absorbing transition.
We also expect the ordinary DP-type scaling in the transient regime, 
starting from a single pair of particles, in contrast to the 
controversial transient behaviors in the PCP model.

At nonzero $r$, the system escapes from the no-pair subspace with finite 
probability. The order parameter can be significantly influenced by the memory
effect induced by this in-and-out dynamics with respect to the no-pair subspace. 
The parameter $r$ connects the DP fixed point ($r=0$) to the PCPD point ($r=1$).
Therefore, the GPCPD allows us to study systematically the origin of 
the new type of critical behavior found in the PCPD. 

\begin{figure}
\includegraphics*[width=.9\columnwidth]{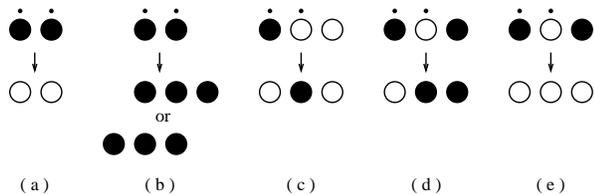}
\caption{Illustration of dynamic rules. Filled~(Empty) circles
represent occupied~(empty) sites. Dots indicate selected sites.}
\label{fig:gpcpd}
\end{figure}
Dynamics of the GPCPD can be implemented in Monte Carlo simulations as
follows~(see Fig.~\ref{fig:gpcpd}). 
First, select a pair of sites $(i,i+1)$ at random.
When  both  sites are occupied, the two particles (a) annihilate with
probability $(1-d)p$ or (b) branch a particle at one of the neighboring sites,
$i-1$ or $i+2$, with  probability $(1-d)(1-p)$.
Branching attempt to an already occupied site is rejected.
When only one site is occupied, (c) the particle hops to the other site
with probability $d$. If the hopping (not branching) creates a new  particle pair,
the pair (d) survives with probability $r$ or (e) annihilates
with probability $1-r$. When both sites are empty, no change is made.
The time increases by one unit after $L$ such trials. 

To speed up simulations, we adopted a technique utilizing a list of active
pairs of neighboring sites. A neighboring site pair is stored in the active pair list
if it contains at least one particle. Then, a site pair in the list is selected
randomly for dynamics. The time increases by 
$1/N_{\rm pair}$ with $N_{\rm pair}$ the number of active pairs in the list 
at each attempt.

\section{Monte Carlo simulations}\label{sec:MC}
Monte Carlo simulations were performed to investigate critical behaviors of
the GPCPD. The critical points are approached by varying $p$ at each 
$r=0$, 0.25, 0.5, 0.75, and 1 with fixed diffusion probability $d=0.1$.

\subsection{Defect simulations}\label{subsec:def} 

We performed the so-called defect simulations to locate the critical points.
Starting with a single pair of particles, we measured the survival
probability $P(t)$ that the system is {\em surviving} at time $t$, the
number of particle pairs $N(t)$ averaged over all samples, and 
the mean distance of spreading $R(t)$ averaged over the surviving samples.

Our definition of surviving samples is different from the
conventional one where all samples not being trapped into one of 
the absorbing states are considered as surviving ones. Here, 
only samples with at least one particle pair are regarded as surviving. 
Samples with only solitary diffusing particles (no pair but irrespective of the number
of particles) are regarded as {\em dead} ones, even if they are not trapped completely.
They may be dead for a while, but can resurrect later 
as surviving samples when diffusing particles meet and form 
a particle pair.  The dead states form the no-pair subspace (see Sec.~\ref{sec:model}).
At nonzero $r$, the system can evolve in and out of the no-pair subspace. 
With this definition, the survival probability $P(t)$ actually 
represents the probability that the system contains at least one particle pair,
or equivalently the probability that the system stays outside the no-pair subspace.

The conventional surviving ensemble includes the most states inside the no-pair
subspace except the {\em true} absorbing states (the vacuum state and the states 
with a single particle). The survival probability is now dominated by the trivial pair 
annihilation dynamics of diffusing particles inside the no-pair subspace \cite{dressing},
which does not reflect the proper dynamic and stationary scaling behavior of the order parameter 
(the pair density). Hence, the fluctuations of the order parameter should be 
described in our surviving ensemble defined as the complement of the no-pair subspace.

At criticality, the values of the measured quantities decay algebraically as~\cite{Grassberger79}
\begin{eqnarray}\label{dyn-exp}
P(t) &\sim t^{-\delta'}, \nonumber\\
N(t) &\sim t^{\eta}, \\
R(t) &\sim t^{1/z}, \nonumber
\end{eqnarray}
and the double logarithmic plots against time show straight lines.
Off criticality, these plots show curvatures in the long time limit.
Precise estimates for the critical points and the scaling exponents 
can be obtained by examining the local slopes of the curves.
The local slope, called as the effective exponent, is defined as
$$
\delta'(t) = - \frac{ \ln[ P(t) / P(t/m) ] }{\ln m}
$$
and similarly for $\eta(t)$ and $z(t)$ with a constant $m$~(taken to be 10). 
The power-law scaling behavior implies that they converge to the values
of $\delta'$, $\eta$, and $z$ asymptotically as $t\rightarrow\infty$ at
$p=p_c$. At off-critical points, they behave like being at the critical point
in the early time regime and then deviate to a trivial value characteristic of 
the active or inactive phase. From this crossover behavior one can
determine the location of the critical point.

\begin{figure}
\includegraphics*[width=\columnwidth]{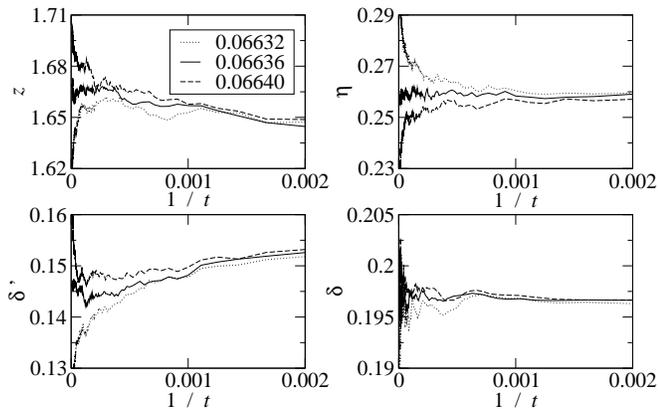}
\caption{The effective exponents for the GPCPD at $r=0.5$. 
}\label{fig:defR50}
\end{figure}
The defect simulations were performed up to $10^5$ time steps and
the observables were measured and averaged over $\sim 2\times 10^5$ samples.
Figure~\ref{fig:defR50} shows plots of the effective exponents against $1/t$
at $r=0.5$. Apparently the effective exponents for $\delta'$, $\eta$, and $z$ 
converge to their asymptotic values at $p=0.06636$, while they 
clearly bend up or down with a curvature at $p=0.06632$ and $0.06640$.
It leads us to estimate that $p_c = 0.06636(4)$ and
$z=1.67(3)$, $\eta=0.26(1)$, $\delta'=0.14(1)$. 

The errors in the exponent values mainly stem from the uncertainty
in the $p_c$ estimate. Statistical errors are much smaller than this
systematic error, which can be clearly seen in Fig.~\ref{fig:defR50}.
A correction to the scaling could 
lead to a systematic error in the estimate of $p_c$, and hence, of critical
exponents. The plots in Fig.~\ref{fig:defR50} show that
the correction is quite small for $N(t)$.  The effective exponents $\delta'(t)$
and $z(t)$ have a little time dependence~(seemingly, 
linear dependence in $1/t$) even at the estimated $p_c$, 
while such time dependence is negligible for $\eta(t)$ at the
estimated $p_c$. So the best estimate for $p_c$ is obtained from 
the plot of $\eta(t)$ vs. $1/t$~\cite{comment}. 
We also find similar behaviors for other values of $r$ except $r=1$.

We also plot the combination of the effective exponents 
$\delta(t) \equiv 1/z(t)-(\eta (t)+\delta' (t))$. 
At criticality, this exponent should converge asymptotically to
the ratio of two stationary scaling exponents, $\beta/\nu_\parallel$
(known as the initial slip exponent), if the hyperscaling relation~\cite{Mendes} is satisfied. 
At $r=0.5$, we obtain that $\delta=0.197(5)$. This value will be compared to 
$\beta/\nu_\parallel$ estimated independently later in the static Monte Carlo simulations. 
The same analysis is repeated for other values of
$r$ and the results are summarized in Table~\ref{table:def}. 

\begin{table}
\caption{Critical points and critical exponents estimated from the defect simulations.
As a reference, corresponding values of the DP and DI classes, taken from
Ref.~\cite{Jensen94}, are given.}
\label{table:def}
\begin{ruledtabular}
\begin{tabular}{c|cccccc}
$r$  & $p_c$ & $z$ & $\eta$ & $\delta'$ & $\delta$ 
\\ \hline
0    & 0.04687(2) & 1.58(1) & 0.314(6) & 0.160(5) & 0.160(5) \\
0.25 & 0.05505(5) & 1.62(3) & 0.29(1) & 0.15(1) & 0.175(5) \\
0.50 & 0.06636(4) & 1.67(3) & 0.26(1) & 0.14(1) & 0.197(5) \\
0.75 & 0.08315(5) & 1.75(5) & 0.20(2) & 0.13(2) & 0.235(5) \\ 
1    & 0.1112(1)  & 1.7(1)  & 0.18(5) & 0.09(2) & 0.31  \\ \hline \hline
DP   &         & 1.579(2) & 0.3137(10) & 0.1596(4) & 0.160(2) \\ \hline
DI   &         & 1.753(3) & 0.000(1) & 0.285(2) &  0.285(4)  \\ 
\end{tabular}
\end{ruledtabular}
\end{table}

The two exponents $\delta$ and $\delta'$ do not necessarily coincide unless 
the evolution operator is invariant under the time-reversal transformation~\cite{Haye}.
However, it is well known in models with multiple absorbing states that
these two exponents coincide if one starts with the so-called 
{\em natural} initial configurations in defect simulations. 
This aspect will be discussed in details elsewhere~\cite{unpub1}.

\begin{figure}
\includegraphics*[width=\columnwidth]{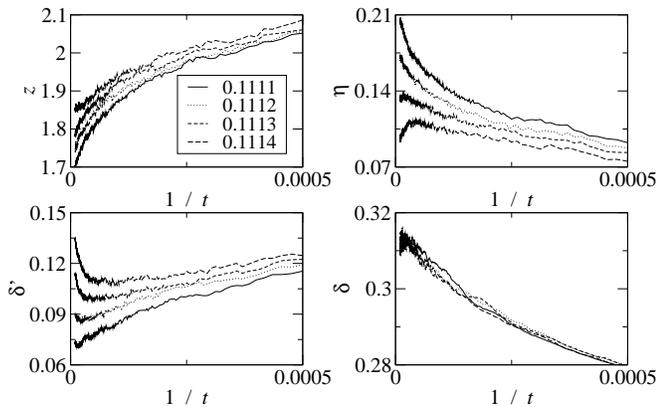}
\caption{The effective exponents for the PCPD~($r=1$).}\label{fig:defR100}
\end{figure}

The PCPD point at $r=1$ is an exceptional case. 
Very strong corrections to scaling are observed in all quantities 
as can be seen in Fig.~\ref{fig:defR100}. 
It has been already noted in Refs.~\cite{Haye2,Odor1}. 
We tried to locate the critical point from the curvature change in the
plots of the effective exponent vs.~$1/t$. The plot of $\eta(t)$ 
vs.~$1/t$ shows that $p_c\simeq 0.1113$, while the plot of $\delta'(t)$ vs.
$1/t$ does $p_c\simeq 0.1112$. It suggests that the asymptotic scaling 
regime has not been reached yet until $10^5$ time steps at the PCPD point.
The exponent $\delta$ shows the worst behavior. 
Even one can not see any difference between super-critical and sub-critical
behaviors~(so we could not estimate an error bar of $\delta$). 
So the values of $p_c$ and critical exponents at
the PCPD limit contain the largest uncertainty. 
It may be quite interesting to investigate the origin and the nature of the
unexpected huge corrections at the PCPD point, but we do not have
any reasonable explanation at this moment.

When one faces a strong correction-to-scaling behavior in investigating
numerically unknown critical phenomena of a model system, one may 
perform an extensive simulation up to much larger length scales
and much longer time scales. However, if it is beyond the present
day's computing capability, one should look for an efficient alternative model
which presumably exhibits the same critical behavior with smaller
correction to scaling. The PCPD model applies to such a case. 

The GPCPD model may serve as an efficient alternative model for the PCPD 
model. The $r=1$ (PCPD) point of the GPCPD model is not a {\em special point}; 
it possesses neither an additional symmetry nor  conservation law,
compared to other points for $0<r<1$. Only the strength of the long-term memory
effect changes with $r$. One may guess that the GPCPD model with  all nonzero
$r$ may belong to the same universality class as the PCPD model and
hope that the corrections to scaling are controllable for small $r$.
It turns out that the GPCPD model possesses much smaller corrections to scaling
for $r<1$ (at least up to $r=0.75$). So the critical points are very accurately estimated
and the critical exponents are determined with reasonable errors.

At $r=0$, the values of all exponents are in excellent accord with the DP values.
It confirms the expectation that the GPCPD without the memory effect 
should belong to the DP class. For other values
of $r\neq 0$, the exponent values begin to deviate from the DP values. 
They are also clearly different from the DI values. 
It confirms that the GPCPD with
the long-term memory effect~($r\neq0$) displays novel type critical phenomena that
do not belong to the DP nor DI universality class.
More importantly and very surprisingly, the exponent values show a dependence on $r$
(see Table~\ref{table:def}). This opens up a new possibility
that the novel universality class would be characterized by
continuously-varying critical exponents, i.e.,~not a fixed point
but a fixed line parametrized by the memory strength $r$. 
This remarkable finding should be examined carefully 
whether it is also present in the scaling property of the steady states. 

\subsection{Static simulations}\label{subsec:static}

The criticality in the steady states is studied via the so-called static Monte Carlo
simulations where one starts with a macroscopically occupied configuration
on finite size lattices. Here, we start with the fully occupied configuration
and use the periodic boundary conditions.
As the order parameter, we measure the pair density $\rho_s (t)$
averaged over surviving samples that contain at least one particle pair
at a given time $t$, or equivalently averaged over the complementary states to the
no-pair subspace, as defined in the defect simulations previously. 
The averaged quantities over the conventional surviving ensemble would also describe 
the system properly in the early time regime, up to the time scale when the system 
typically enters into the no-pair subspace. However, since then, the trivial pair 
annihilation dynamics of diffusing particles governs the system evolution \cite{dressing} 
and the average value of the order parameter crosses over to the trivial stationary value.

The pair density averaged over our surviving ensemble satisfies a finite-size-scaling form as 
\begin{equation}\label{fss0}
\rho_s(\varepsilon,t,L) = L^{-\beta/\nu_\perp} f(\varepsilon
L^{1/\nu_\perp}, t  / L^z) \ ,
\end{equation}
where $L$ is the system size and 
$\varepsilon \equiv p_c-p$ is the distance from the critical point $p_c$.
The exponents $\beta$ and $\nu_\perp$ are the critical exponents associated with
the order parameter and the spatial correlation length, respectively.
The dynamic exponent $z$ is the ratio of the two exponents; $z=\nu_\parallel/\nu_\perp$ 
where $\nu_\parallel$ is the relaxation time exponent~(see Ref.~\cite{Haye} for a review).

At criticality~($p=p_c$), we have
\begin{equation}\label{rho_t}
\rho_s(0,t,L) = L^{-\beta/\nu_\perp} g(t/L^z) \ ,
\end{equation}
where  $g(x)\sim x^{-\beta/\nu_\parallel}$ for small $x$
and becomes a constant for large $x$.
Thus, at $p=p_c$,  
the pair density decays as $\rho_s \sim t^{-\beta/\nu_\parallel}$ for 
$t\ll L^z$ and saturates to a steady-state value  
$\rho_s \sim L^{-\beta/\nu_\perp}$ for $t\gg L^z$. 

The power-law scaling behavior in the transient regime~($t\ll L^z$) 
can be studied with an effective exponent 
$$[\beta/\nu_\parallel] \equiv -\frac{\ln [\rho_s(t)/\rho_s(t/m)]}{\ln m}$$ 
with a constant $m$.
It converges to $\beta/\nu_\parallel$
for large $t$ ($\ll L^z$) at the critical point $p=p_c$, but
deviates from it at $p\neq p_c$. Using this property we could determine
the critical point $p_c$ independently and hence the exponent 
$\beta/\nu_\parallel$.

\begin{figure}
\includegraphics*[width=\columnwidth]{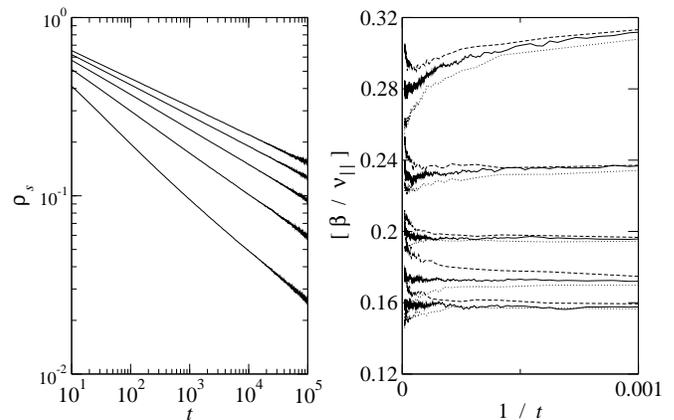}
\caption{{\bf Left}: Order parameter decay at $p=p_c$ at each
$r=0.0$ (top),$\ldots,1.0$ (bottom). {\bf Right}:
The effective exponent $[\beta/\nu_\parallel]$ defined with $m=8$.
Solid lines represent the effective exponents at the estimated critical
points~(see text) and dotted~(broken) lines at  
super-critical~(sub-critical) points at each $r=0.0$
(bottom),$\ldots,$1.0 (top).}\label{dynam-exp}
\end{figure}

The order parameter is measured in a lattice of size $L=10^4$ 
up to $t=10^5$ time steps and averaged over $2000\sim 5000$ samples.
The finite-size saturation effect is invisible up to $t=10^5$ for this system size.
Figure~\ref{dynam-exp} shows the log-log plots of the order parameter at
the estimated critical points as well as the plots of the effective exponents 
at and near the criticality. From these plots, we estimate  
$p_c = 0.04687(2)$, $0.05505(5)$, 
$0.06637(2)$, $0.08317(4)$, and $0.1113(1)$ at each value of $r=0$, $0.25$, 
$0.5$, $0.75$, and $1$, respectively.
Except for the $r=1$ case, the values of the effective exponents at the estimated 
$p_c$ do not show any significant systematic error and seem to reach their
asymptotic values quite early, at around $t=10^3$. Statistical fluctuations
are also very small. The off-critical data in Fig.~\ref{dynam-exp} at the
values of $p$ displaced by the amount of error bars in the above $p_c$ estimates,
clearly show either upward or downward curvatures, which demonstrates 
the accuracy of our $p_c$ estimate. 
Moreover, these $p_c$ values are in very good agreement with those 
obtained from the defect simulations (see Table~\ref{table:def}).

The values of $\beta/\nu_\parallel$ are given by the limiting values of the
effective exponents for large $t$ at the estimated $p_c$, which
are presented in Table~\ref{table:static}. For consistency, 
we use the defect simulation results for the $p_c$ values and their errors.
The estimated values of $\beta/\nu_\parallel$ are again in excellent
accord with the values of $\delta=1/z-(\eta+\delta')$ measured in the defect simulations
for $r\neq 1$, which implies that the hyperscaling relation~\cite{Mendes}
holds in this model. 

Note that the correction to scaling is significant at the PCPD point~($r=1$),
as found in the defect simulations. At $r=1$, the effective exponent plot shows
a strong time dependence even at the estimated critical point and there is a noticeable
curvature in Fig.~\ref{dynam-exp}.
It leads to a rather large error in the $p_c$ estimate and, hence, 
the exponent estimates. Our estimate of $\beta/\nu_\parallel$  also
shows a noticeable discrepancy with the estimate of $\delta$ 
(see Tables~\ref{table:def} and \ref{table:static}). However, 
this discrepancy presumably comes from the inaccurate estimation
of $p_c$ at $r=1$ and insufficient time steps getting into the asymptotic regime.
So it cannot be regarded as an evidence for the breakdown of the
hyperscaling relation at the PCPD point.

\begin{table}
\caption{Critical exponents of the GPCPD and corresponding values of the DP
and DI classes, taken from Ref.~\cite{Jensen94}.}
\label{table:static}
\begin{ruledtabular}
\begin{tabular}{c|cccccc}
$r$  & $\beta/\nu_\parallel$ & $\beta/\nu_\perp$ & $z$ & $\nu_\perp$ &
$\beta$ \\ \hline
0   &0.159(1)&0.252(3) &1.58(1)&1.10(1)&0.277(5)\\ 
0.25&0.173(5)&0.282(5) &1.64(5)&1.10(3)&0.310(14)\\
0.50&0.197(3)&0.330(6) &1.69(5)&1.10(3)&0.363(17)\\
0.75&0.230(5)&0.40(1)  &1.72(5)&1.17(5)&0.468(30)\\
1   &0.27(4) &0.50(4)  &1.8(2) &1.30(10)&0.65(12)&\\ \hline \hline
DP  &0.1596(4) & 0.2522(6) & 1.5798(18) & 1.0972(6) & 0.2767(4) \\
DI  &0.285(2) & 0.500(5) & 1.750(5) & 1.84(6) & 0.92(3)  \\ 
\end{tabular}
\end{ruledtabular}
\end{table}

The steady-state pair density for $t\gg L^z$ satisfies
\begin{equation}\label{rho_s}
\rho_s(\varepsilon,t=\infty,L) = L^{-\beta/\nu_\perp}
h(\varepsilon L^{1/\nu_\perp}) \ ,
\end{equation}
where the scaling function behaves as $h(x) \sim x^\beta$ for large $x$ and
becomes a constant near $x\simeq 0$.  At criticality ($\varepsilon =0$),
it decays algebraically with size as $\rho_s \sim L^{-\beta/\nu_\perp}$.
Utilizing this algebraic scaling property, we could also estimate the 
location of the critical points $p_c$. They again agree perfectly well 
with the previous other results.

\begin{figure}
\includegraphics*[width=\columnwidth]{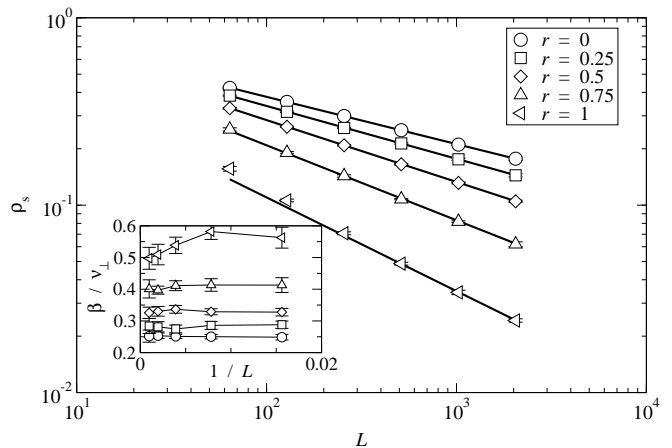}
\caption{The power-law scaling behavior of the order parameter at $p=p_c$ 
at each $r$. The solid line is a guide for the eyes whose slope is 
$\beta/\nu_\perp$ in Table~\ref{table:static}.}\label{p_c}
\end{figure}

We run $10^3\sim 10^4$ samples up to $4\times 10^6$ time steps 
for the system size $L=2^6,\ldots,2^{11}$ at the estimated critical
points given in Table~\ref{table:def}. 
Plots of $\rho_s$ vs.~$L$ at each critical point
are presented in Fig.~\ref{p_c}. 
A least-square straight-line fitting in the log-log plot might yield 
the value of $\beta/\nu_\perp$. However, it would result in an inaccurate
estimate in the presence of a strong correction to scaling. 
Instead, we made use of an effective exponent $\left[ \beta/\nu_\perp \right] =
-\ln [\rho_s(2L)/\rho_s(L)] / \ln 2$ to extract the accurate exponent value
from the extrapolation. 
It is plotted in the inset of Fig.~\ref{p_c} and the results are presented in
Table~\ref{table:static}. As before, there exists a strong correction to
the power-law scaling at the PCPD point at $r=1$, whereas such an effect
is very small for the other values of $r$.

\begin{figure}
\includegraphics*[width=\columnwidth]{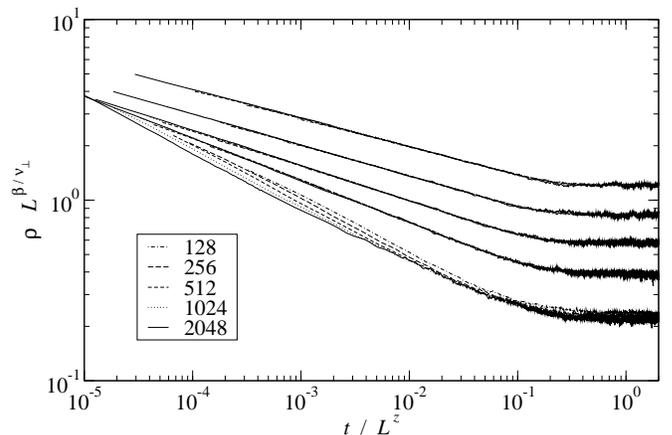}
\caption{Scaling plots according to Eq.~(\ref{rho_t})
at each $r=0.0$ (top),$\ldots,1.0$ (bottom). Each data set 
is shifted vertically by a constant factor to avoid an overlap with others
at different values of $r$.}
\label{d-collapse}
\end{figure}

With $p_c$ and $\beta/\nu_\perp$ determined, we could estimate the dynamic
exponent $z$ using the scaling form in Eq.~(\ref{rho_t}). It was estimated 
as the optimal value that yields the best collapse of 
$\rho_s(\varepsilon=0,t,L)$ for $L=2^7,\ldots,2^{11}$ (see Fig.~\ref{d-collapse}).
The resulting values of $z$ are presented in Table~\ref{table:static}.
They are consistent with the ratio of the two exponent ratios, 
$\beta/\nu_\perp$ and $\beta/\nu_\parallel$, and agree very well 
with the values measured in the defect simulations.

\begin{figure}
\includegraphics*[width=\columnwidth]{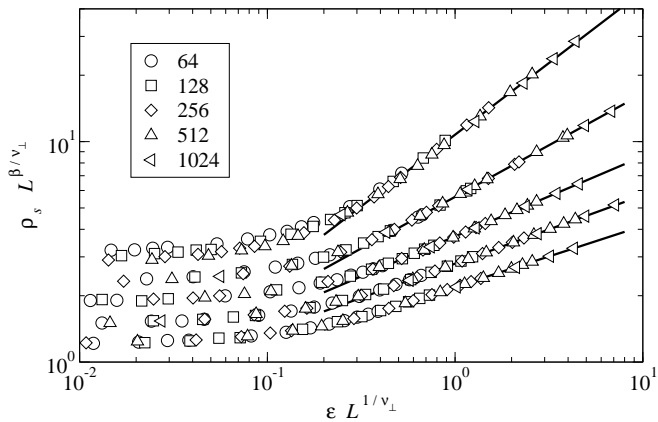}
\caption{Scaling plots according to Eq.~(\ref{rho_s})
at each $r=0.0$ (bottom),$\ldots,1.0$ (top). The solid line is a guide for
the eyes whose slope is $\beta$ in Table~\ref{table:static}.
Each data set is shifted vertically by a constant factor to avoid an
overlap with others at different values of $r$.}
\label{s-collapse}
\end{figure}

We also determined the value of 
$\nu_\perp$ by collapsing off-critical steady-state data of 
$\rho_s(\varepsilon,t=\infty,L)$ in the scaling plot, using Eq.~(\ref{rho_s}).
In Fig.~\ref{s-collapse}, the data for each $r$ collapse well on a 
single curve with a specific value of $\nu_\perp$, respectively.
The order parameter exponent $\beta$ is then obtained by
a product of $\beta/\nu_\perp$ and $\nu_\perp$ and presented in
Table~\ref{table:static}. They are equal to the slopes of the scaling
function $h(x)$ for large $x$ in the log-log scale as shown in 
Fig.~\ref{s-collapse}.

At the PCPD point~($r=1$), our results are consistent with those obtained by
Carlon {\em et al.}~\cite{Carlon}~($p_c = 0.111(2)$, $z=1.87(3)$,
$\beta/\nu_\perp=0.50(3)$) and
Hinrichsen~\cite{Haye2}~($p_c=0.1112(1)$, $\beta/\nu_\parallel=0.25(2)$,
$z=1.83(5)$, $\beta<0.67$). Although some of critical exponent ratios are
close to the DI values, the critical exponents are incompatible with those
of the DI class. They are also inconsistent with the DP universality class.
The strong correction-to-scaling behaviors were observed
at the PCPD point with $r=1$ in both the defect and the static simulations.
So our estimates for the critical exponents contain considerable 
uncertainty at the PCPD point. 
However, such corrections are not prominent for other values of $r$, 
which enables us to estimate the values of the critical exponents 
very accurately for $r<1$. 
As expected, the values of the critical exponents at $r=0$ 
agree very well with the DP values.

For $0<r<1$, the correlation length exponent $\nu_\perp$ seems to remain
the same as the DP value for a wide range of $r$, at least up to 0.75.
The value of the relaxation
time exponent $\nu_\parallel= z \nu_\perp $ seems to increase slightly with
increasing $r$ (10\% up to $r=0.75$),  but, within present numerical accuracy,
it may be difficult to conclude that its variation  is real and 
not due to corrections to scaling.
However, it is clearly visible that the order parameter exponent 
$\beta$ (also $\beta/\nu_\perp$ and $\beta/\nu_\parallel$)
varies considerably with $r$ (more than 60\% !).
It clearly signals a novel universality class with scaling exponents varying 
continuously with the parameter $r$. 

Summing up all results, our Monte Carlo simulations show 
that the GPCPD displays the novel critical phenomena distinct from the DP and DI classes, 
with  continuously varying exponents depending on the strength of
the long-term memory effect $r$.
These results are quite surprising and remarkable. 
In the renormalization group language, it implies that 
there is a fixed line parametrized by $r$, instead of  fixed points at 
two end points; the DP ($r=0$) and the PCPD point ($r=1$).

Continuously varying critical exponents in non-equilibrium systems
are very rare. Systems with infinitely many absorbing states like
the PCP may display continuously varying exponents depending on
initial conditions, but only the exponents describing 
non-stationary properties ($\delta'$ and $\eta$) are
varying~\cite{PCP}.  This variance is even disputed very
recently~\cite{GCR,DH}.
L{\'e}vy-type long-range flights are known to
be relevant to absorbing critical phenomena and the stationary
critical exponents vary continuously with the exponent describing
the long-range tail distribution of the
flights~\cite{Jan_Hin}. However, the GPCPD involves only
short-range processes. Recently, multi-species particle
reaction-diffusion systems with inter-species hardcore
interactions are conjectured to display continuously varying
stationary-state exponents, but the transition occurs at the
trivial annihilation point~\cite{Kwon}. Moreover, the variance is
very small and it would be extremely difficult to confirm it numerically.
Our studies suggest strongly that the GPCPD belongs to a
completely new type of universality class with continuously varying
exponents which has not been explored before.

\section{Universality}\label{sec:universality}

In order to establish the novel universality class firmly,
we study other model systems that share the common feature of the memory
effect with the GPCPD. As described in Ref.~\cite{Haye3},
the PCPD can be regarded in a coarse-grained level as a cyclically coupled 
system of two particle species:
one species ($A$) performing DP-like dynamics and the other species ($B$)
the annihilating random walks.  
In the GPCPD model, a particle pair corresponds to an $A$  particle
and a solitary particle to a $B$ particle.
Two species are coupled through transmutations, which leads to the long-term 
memory effect.
We set up two different 1D models of a cyclically coupled system
and investigate numerically their scaling behavior to check 
our novel universality class. 
The two typical DP dynamics are employed for the $A$-particle dynamics,
i.e.~the branching-annihilating random walk model with one 
offspring (BAW1)~\cite{BAW} and the contact process. 

The first model (referred to as the ABB model) adopts the BAW1 evolution rule 
for the $A$ dynamics. Each lattice site is either occupied by an $A$ or $B$ particle, 
or empty. First, select a site at random. When it is occupied with an $A$,
the $A$ particle (a) branches an $A$ with probability $(1-p)$ on a neighboring
site, or (b) hops to a neighboring site with probability $p(1-\mu)$, 
or (c) transmutes to a $B$ particle with probability $p\mu$.
When the branched or hopping $A$ particle lands on another $A$ particle,
both $A$ particles annihilate immediately as in the BAW models.
In case that it would land on a $B$ particle, the trial is rejected and
the $B$ particle transmutes to the $A$ particle.
When the selected site is occupied with a $B$ particle,
the $B$ particle hops to a neighboring site with probability $d$ or
does nothing with probability $1-d$.
When it would land on another particle (either $A$ or $B$), the trial is rejected
and all involved $B$ particles transmute to $A$ particles.
When the selected site is vacant, nothing happens. 

The evolution rule of the second model (the ABC model) is
almost equivalent to the ABB model, except that the contact process
is adopted for the $A$ particle dynamics. An $A$ particle 
(a) branches an $A$ with probability $(1-p)$ on a neighboring
site, or (b) vanishes spontaneously with probability $p(1-\mu)$, 
or (c) transmutes to a $B$ particle with probability $p\mu$.
When the branched $A$ particle would land on either $A$ or $B$ particle,
the trial is rejected and the involved $B$ particle transmutes to the $A$ particle.
A $B$ particle hops to a neighboring site with probability $d$ or does
nothing with probability $1-d$, as in the ABB model.
When it would land on another particle (either $A$ or $B$), the trial is rejected
and all involved $B$ particles transmute to $A$ particles.

In both models, the two species of particles are coupled cyclically through the 
transmutations. The transmutation rate from $A$ to $B$ is controlled explicitly 
by the parameter $\mu$. As $\mu$ increases, the transmutation events 
occur more likely . At $\mu=0$, the $A\rightarrow B$ channel is completely 
blocked. Therefore there is no feedback mechanism to change the $A$ particle
density through intermediate $B$ particles. 
The transmutation rate from $B$ to $A$ is implicit and
determined by the evolution rule and the other control parameters. 
The $B\rightarrow A$ channel relies heavily on the the diffusive property of 
$B$ particles. With nonzero $d$, the $B$ particles diffuse until they meet 
another particle and transmute to the $A$ particles. This process invokes
the same mechanism that gives rise to the long-term memory effect
in the GPCPD model for $\mu\neq 0$. The $B$ particles created by the
$A$ particles through transmutations perform the random walks
before transmuting back to the $A$ particles at later times.
At $d=0$, the $B\rightarrow A$ channel is still open, but only with the
short-term memory processes, which should be irrelevant.

\begin{figure}
\includegraphics*[width=\columnwidth]{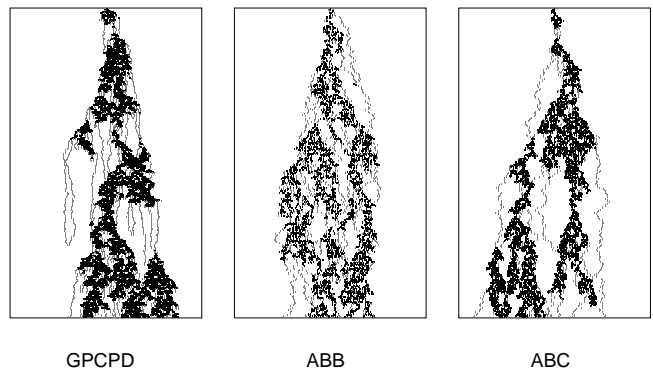}
\caption{Critical spreading of activities originated from a 
seed~(a particle pair for
the GPCPD and an $A$ particle for the ABB and ABC models). A black
pixel represents a particle pair or an $A$ particle, and a gray one a
solitary particle or a $B$ particle. The horizontal~(vertical~(down)) 
direction corresponds to the space~(time) direction.}
\label{fig:conf}
\end{figure}

We expect to observe the same type of critical behaviors,
as in the GPCPD model, characterized by the continuously varying critical
exponents. We performed the defect and static simulations to locate
the critical points and estimate the critical exponents of the ABB and
the ABC models at several values of $\mu$ with fixed hopping 
probability $d=1$. In Fig.~\ref{fig:conf}, we compare the space-time 
structures of spreading patterns of activities in the GPCPD~($r=0.5$), 
ABB~($\mu=0.5$), and ABC~($\mu=0.2$) models at critical points. 
In all cases, the intermediate long-range diffusive processes of 
the solitary particles (or the $B$ particles) are commonly observed
between the particle pairs (or $A$ particles) at different space-time position.
The time scale of these processes represents the effective {\em life time} 
of the intermediate $B$ particles, that seems to be comparable to
the simulation time. This aspect will be discussed quantitatively in the
next section.

\begin{figure}
\includegraphics*[width=\columnwidth]{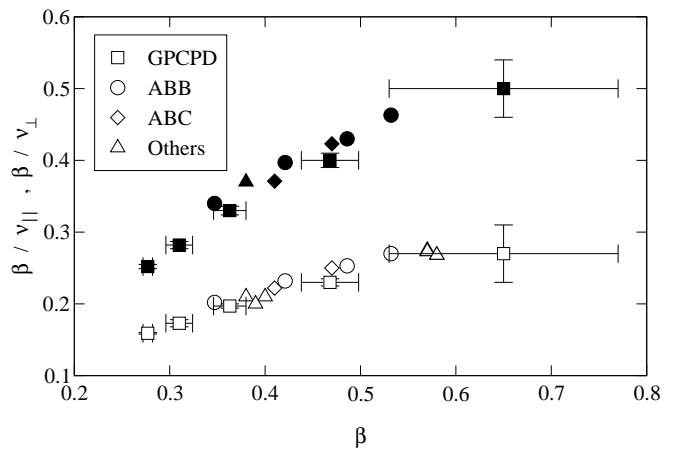}
\caption{Parametric plots of $\beta/\nu_\perp$ (filled symbol) and
$\beta/\nu_\parallel$ (open symbol) with respect to $\beta$ of the GPCPD,
the ABB, and the ABC models. The data taken
from Refs.~\cite{Haye2,Odor1,Haye3,KPark,Odor2} are also plotted and 
labelled by``Others".}
\label{fig:exp}
\end{figure}

We take the $A$ particle density as the order parameter and use the
same definition of the surviving samples as in the GPCPD 
model, i.e.~only the samples with at least one $A$ particle are regarded as
surviving ones. Here, we only state the results without showing the data.
Our numerical simulations confirm that both $ABB$ and $ABC$ models
show continuously varying exponents with the parameter $\mu$.
As expected, both models converge to the DP class at $\mu=0$.

In Fig.~\ref{fig:exp}, the values of the critical exponents at various 
transmutation rates are plotted together with those for the GPCPD model
and those available from previous other 
works~\cite{Haye2,Odor1,Haye3,KPark,Odor2}.
Remarkably, they are lying along a single smooth line.
It indicates that all these models fall into the same novel universality class
with continuously varying exponents that can be parametrized by a single
parameter (memory strength). 
The lines for $\beta/\nu_\perp$ and $\beta/\nu_\parallel$ seem to be 
almost linear for small $\beta < 0.5 $. It implies that the values of 
the correlation exponents, $\nu_\perp$ and $\nu_\parallel$, do not vary 
too much from the DP values, in contrast to a wide variation of the 
order parameter exponent $\beta$.

\section{Life time distribution of intermediate particles}\label{sec:lifetime}

The continuously varying exponents of single parameter family
remind us of the equilibrium Gaussian model in two dimensions
where a marginal operator accounts for the continuous variation of
the critical exponents. Here, we test a similar scenario that
the long-term memory effect mediated by the solitary particles $(B)$ may
play the role of the marginal operator.

In a heuristic way, one can write down a mean field type
rate equation for the order parameter, the pair ($A$) density $\rho(t)$, as
\begin{equation}\label{rate_eq}
\frac{d\rho(t)}{dt} = a \rho(t)  - b \rho(t)^2 +
\lambda \int^{t}_0 dt' F(t-t') \rho(t') \ ,
\end{equation}
where the first and the second terms appear generically representing the
spontaneous annihilation and branching of the pairs $(A)$.
The $\lambda$ term represents the transmutation of the solitary 
particles~($B$), which were detached (created) from $A$ at time $t'$,
into $A$ at later time $t>t'$. Therefore, the kernel $F(\tau)$ represents the
distribution function of the life time $\tau$ of the $B$ particles.

\begin{figure}
\includegraphics*[width=\columnwidth]{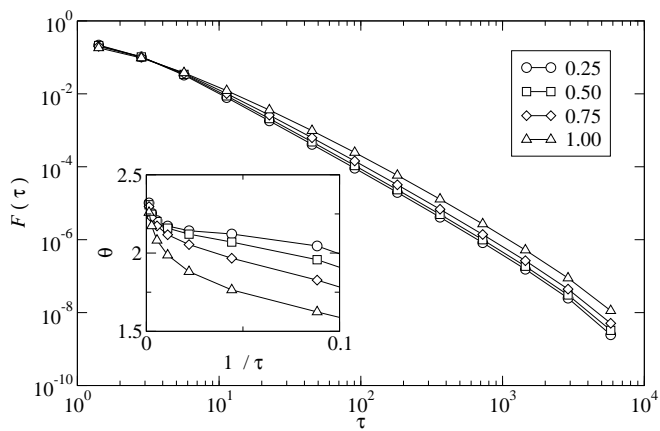}
\caption{Log-binned life time distribution of solitary particles at the
critical points of the GPCPD with $r=0.25,\ldots,1.00$.
{\bf Inset}: Effective exponent for $F(\tau)\sim \tau^{-\theta}$.
The apparent curvature for large $\tau$ $ > 10^3$ is simply due to the finite 
simulation time of $t_{\rm max}=10^4$.}
\label{fig:tau}
\end{figure}

We expect that $F(\tau)$ should have a power-law type distribution, 
because the transmutation event from $B$ to $A$ is governed by an annihilating 
random walk process of two $B$ particles. 
We measure numerically $F(\tau)$ at the critical point from the histogram of
life time $\tau$ of solitary ($B$) particles.
We start with a single $A$ on an infinite lattice,
and generate $10^5$ samples up to $t_{\rm max}=10^4$ to obtain at least $10^9$ 
$B$'s. Figure~\ref{fig:tau} shows $F(\tau)$ averaged over surviving samples 
for the GPCPD model.  It follows a power law
\begin{equation}
F(\tau) \sim \tau^{-\theta} \ .
\end{equation}
The values of $\theta$ are determined from the effective exponent plots 
as shown in the inset. 
After the transient regime at small $\tau$, the effective exponent
converges to $\theta =2.25(5)$ for all values of $r$.

The apparent deviation from the power-law distribution 
for large $\tau>10^3$ is simply due to the finite simulation time of
$t_{\rm max} =10^4$. The number of $B$ particles with a long lifetime
comparable to $t_{\rm max}$ must have been underestimated in our simulations, 
which leads to a slight downward deviation of $F(\tau)$ for large $\tau$. 
Therefore, the apparent blowup of the effective exponent $\theta$
near $1/\tau=0$ in the inset of Fig.~\ref{fig:tau} should be ignored. 
The power law distribution with $\theta=2.25(5)$ is observed 
universally for all three models considered in this paper.
Note that the power-law distribution sets in later as one approaches
the PCPD point at $r=1$. 
Presumably this is related to the strong correction-to-scaling
behavior observed in the PCPD.

A straightforward power counting analysis of the mean field 
equation~(\ref{rate_eq}) shows that the renormalization group (RG) scaling
exponent of the $\lambda$ term is given by $y_\lambda = z(2-\theta)$ with
$z$ the dynamic exponent. 
With $\theta>2$, the long-term memory effect appears to be irrelevant
and our scenario for the marginal operator may be discarded.
However, this naive mean-field-type analysis may go wrong 
in the interacting theory with fluctuations considered. 
The L\'{e}vy-flight DP system is such an example~\cite{Jan_Hin}, 
where it is analytically shown by the $\epsilon$-expansion type RG analysis
that the relevancy of long-range flights sets in a little bit earlier than
expected from the mean field theory. We suspect that a similar situation also 
occurs in the GPCPD and the marginal value of $\theta$ becomes slightly bigger than 2.
In this point of view, our scenario may be still alive and even considered 
reasonable because $\theta=2.25(5)$ is not far away from 2.
A definitive test on our scenario needs accurate
determination of the marginal value in a full field-theoretical context, 
which is beyond the scope of the current paper.

\section{Summary and Discussion}\label{sec:sum&dis}

In summary, we introduced a generalized version of the PCPD (GPCPD)
with a parameter controlling the long-term memory effect. 
The GPCPD connects the DP fixed point to the PCPD point continuously.
We investigated numerically the nature of the absorbing phase
transitions for the GPCPD in one dimension. 
Our numerical results strongly suggest that
the GPCPD belongs to the novel universality class,
which is characterized by the long-term memory effect and
the continuously varying critical exponents.
This model can be viewed as the cyclically coupled systems
based on the DP dynamics with the long-term memory effect.
We showed numerically that the two other variants of these systems 
fall into the same universality class. 

We presented one possible scenario to account for this novel universality class
that the intermediate particles generate the long-term memory effect
which may play the role of the marginal operator to the DP fixed point.
We measured the lifetime distribution of the intermediate particles, which
turned out to follow the power law with the exponent value slightly larger than
the marginal value in a naive noninteracting theory.  This small difference 
may disappear in a full interacting theory and the long-term memory effect
may become really marginal. However, this is only speculative and definitely 
needs a full field-theoretical treatment. 

Very recently, there have appeared several works on the PCPD and related
models. Park and Kim~\cite{Park&Kim02} studied three different variants 
of the PCPD along a special line in the parameter space of the diffusion 
and the reaction rates. They found numerically that the exponent values 
for all three models are consistent within statistical errors,
e.g., $\beta/\nu_\parallel \simeq 0.24(1)$. However, this does not imply that
the PCPD scaling with general parameter values can be also described by 
the same fixed point. 

Dickman and de Menezes~\cite{Dickman02} studied the PCPD
using the same ensemble for the surviving samples as we used here.
They found that this ensemble (called as the reactive sector)
is quite useful to study the finite size scaling of the order parameter.
They observed numerically that the critical exponents vary with the 
diffusion rate $d$. Similar to our results, the correlation exponent $\nu_\perp$ 
appears to be independent of $d$ and the same as the DP value.
The relaxation time exponent $\nu_\parallel$ and the order parameter exponent
$\beta$ vary with $d$ up to only about $20\%$ (much smaller than the variation in the 
GPCPD $\sim 60\%$), which leads to an indecisive conclusion.

More recently, Hinrichsen~\cite{Hinrichsen02} introduced a cellular automaton, which
presumably belongs to the same universality class as the PCPD.
Using a parallel update, he simulated the model with size up to
$L=2^{21}$ until $t=2.5\times 10^6$. This work confirms again the difficulty in
studying the critical behavior of the PCPD. The system does not reach the
asymptotic scaling regime at that time scale. From the temporal trend of
the effective exponents, he suggested an extremely slow crossover
to DP, but this suggestion is also far from being conclusive.

Kockelkoren and Chat\'e~\cite{Kockelkoren02} studied general
reaction-diffusion processes without the fermionic constraint. From Monte
Carlo study up to $L=2^{22}$ and $t\sim 10^7$, they obtained
$\beta/\nu_\parallel \simeq 0.200$ in a model characterized with
$XX\rightarrow XXX$ and $XX\rightarrow \emptyset$. They reported
that the same exponents are observed in other similar models, and hence 
claimed that there is a single universality class for the PCPD. 
However, their study does not cover the general cases. 
\'Odor~\cite{Odor02} studied the PCPD with Monte Carlo 
simulations up to $L=10^5$ and $t=10^{8}$, and obtained 
$\beta/\nu_\parallel \simeq 0.21$ for high diffusion rates and
$\beta/\nu_\parallel \simeq 0.25$ for low diffusion rates.
On the other hand, assuming a logarithmic correction, \'Odor obtained
$\beta/\nu_\parallel \simeq 0.21$ also for the low diffusion rates.

These numerical results seem to favor a single non-DP universality class 
for the PCPD. However, as hinted in the work of Dickman and 
de Menezes~\cite{Dickman02}, 
there may be, if any, a rather small variation ($< 20\%$) of the scaling exponents
by changing diffusion rates. So it is not surprising to see
an apparent single universality under the effect of huge corrections to scaling.
In contrast, the exponent variations in the GPCPD are much bigger 
($\sim 60\%$), so it is easier to confirm their variations. 

We believe that  the long-term memory effect  is responsible for unusually long 
relaxation and non-DP scaling, and possibly for continuously varying exponents.  
The long-term memory is also present in the ordinary PCPD model, but it
is controlled implicitly via diffusion rates $d$ of isolated or intermediate particles.
So, it may not be clear to predict how the long-term memory emerges with $d$.
In  the GPCPD model, we directly control the strength of the long-term memory effect
by varying the parameter $r$. 
 
The GPCPD model smoothly connects the DP fixed point $(r=0)$ to the PCPD model
$(r=1)$, in a sense that the exponent values change with $r$ continuously and
monotonically, starting from the DP values. In contrast, the PCPD exponent values 
seem to jump from the DP values and decrease slightly with the diffusion rate $d$. 
For example, as soon as we turn on the diffusion process, 
$\beta/\nu_\parallel$ increases abruptly from the DP value of 0.1595 
to $\sim 0.25$ at $d=0.1$ and slightly decreases to $0.20\sim 0.23$
at high $d=0.7\sim 0.8$~\cite{Dickman02,Kockelkoren02,Odor02}. 
Even if one accepts the claim that this variation is not real and $\beta/\nu_\parallel$
converges to $\sim 0.20$ asymptotically at any nonzero $d$ (single universality class),
one can not avoid the fact that corrections to scaling are much bigger at low $d$ than 
at high $d$. This suggests that the $d\rightarrow 0$ limit in the PCPD model should 
encounter an unusual crossover behavior. In the ordinary crossover,
one expects an interference of the DP fixed point on the PCPD universality class
at $d\simeq 0$,  which generates an apparent exponent value between
the DP and the PCPD value. All numerical results simply disagree with this.
The apparent value (0.25) for $\beta/\nu_\parallel$ at low $d$  is higher than both the 
DP (0.16) and the estimated PCPD value (0.20).
Therefore, the $d=0$ limit in the PCPD model is unusually singular,
which makes it impossible to study the PCPD scaling behavior systematically
starting from the DP fixed point. In contrast, our GPCPD model is generically well suited
to a systematic investigation of the PCPD model by controlling the long-term memory effect
directly.

Besides that, there is one important technical point we adopted in this paper 
for analyzing the numerical data. As explained in Sec.~\ref{sec:MC},
we chose the {\em surviving} ensemble as the collection of samples with at least one
particle pair and the order parameter as the particle pair density. 
With the conventional choice for the surviving ensemble (samples
not trapped in one of the absorbing states), we found that 
the order parameter in finite size systems bears two time scales 
(nontrivial relaxation time and trivial pair annihilation time) and does not show
simple scaling collapse with one-variable scaling function 
like in Eq.~(\ref{rho_t})~\cite{unpub1}. Our choice of the surviving sample should 
involve only one time scale and leads to an excellent scaling as shown 
in Fig.\ref{d-collapse}. With this ensemble, it is natural to choose the particle
pair density (rather than the particle density) as the order parameter.

Summing up, we believe that the GPCPD model serves as an efficient 
generalized model to study the PCPD model systematically.
The GPCPD seems to have much less corrections to scaling, compared to
the ordinary PCPD model, which enables us to present accurate numerical
data for the exponent values. In Figs.~\ref{fig:defR50} and \ref{dynam-exp},
one can clearly see that the asymptotic values for the exponents
set in quite early, at around $t=10^3$, except the PCPD case ($r=1$). 
In order to check any possible long time crossover, we performed 
an extra static simulation for the GPCPD at $r=0.25$ 
on a lattice of size $L=10^5$ up to $t=10^7$. 
We found the more accurate critical point at  $p_c = 0.055045(5)$ and 
the exponent $\beta/\nu_\parallel \simeq 0.172(2)$~\cite{unpub1}. 
The result is in excellent accord with our early-time result on a smaller lattice
(see Table \ref{table:def} and \ref{table:static}).
This confirms again that the GPCPD at $r\neq 1$ does not suffer from 
huge corrections to scaling as observed in Refs.~\cite{Hinrichsen02,Odor02}.

\acknowledgments
We thank P.~Grassberger for his critical reading of this manuscript and
valuable comments. We also thank Su-Chan Park for useful discussions.
This work was supported by grant No.~2000-2-11200-002-3 from the Basic
Research Program of KOSEF.


\begin{references}
\bibitem{addG} A binary spreading process like the PCPD was 
first introduced and studied by Grassberger in~\cite{Grassberger82}.
\bibitem{dressing} This trivial dynamics may be "dressed" by the inteference with 
the local clouds of particle pairs. This issue will be discussed elsewhere.
\bibitem{comment} It does not imply that the estimate using the
plot of $\delta'(t)$ or $z(t)$ vs. $1/t$ would yield a different value of
$p_c$. In such cases the upward and downward curvature at offcritical
points guides the estimation, which leads to the same result.
\end{references}
\end{document}